\documentclass[a4paper]{article}
\usepackage{INTERSPEECH2019}

\usepackage{amsmath,graphicx,pgfplots,tikz}
\usepackage{xfrac,bm,amssymb,multirow,algorithm,algorithmic,float}
\usepackage[caption=false]{subfig}
\pgfplotsset{compat=1.14}

\title{Multiple Sound Source Localization with SVD-PHAT}
\name{Fran\c{c}ois Grondin, James Glass\thanks{This work was supported in part by the Toyoto Research Institute and Signify.}}
\address{Computer Science and Artificial Intelligence Laboratory\\
Massachusetts Institute of Technology\\
    Cambridge, MA 02139, USA}
\email{\{fgrondin,glass\}@mit.edu}

\DeclareMathOperator*{\argmax}{arg\,max}
\DeclareMathOperator*{\argmin}{arg\,min}
\DeclareMathOperator{\Tr}{Tr}
\DeclareMathOperator{\atantwo}{atan2}
\newcommand{\Mod}[1]{\ (\mathrm{mod}\ #1)}
\newcommand{\xMapsto}[2][]{\ext@arrow 0599{\Mapstofill@}{#1}{#2}}

\newcommand{\aiMiCx}{-5.0}
\newcommand{\aiMiCy}{0}
\newcommand{\aiMiCz}{0}
\newcommand{\aiMiiCx}{-3.3}
\newcommand{\aiMiiCy}{0}
\newcommand{\aiMiiCz}{0}
\newcommand{\aiMiiiCx}{-1.7}
\newcommand{\aiMiiiCy}{0}
\newcommand{\aiMiiiCz}{0}
\newcommand{\aiMivCx}{0}
\newcommand{\aiMivCy}{0}
\newcommand{\aiMivCz}{0}
\newcommand{\aiMvCx}{1.7}
\newcommand{\aiMvCy}{0}
\newcommand{\aiMvCz}{0}
\newcommand{\aiMviCx}{3.3}
\newcommand{\aiMviCy}{0}
\newcommand{\aiMviCz}{0}
\newcommand{\aiMviiCx}{5.0}
\newcommand{\aiMviiCy}{0}
\newcommand{\aiMviiCz}{0}

\newcommand{\aiiMiCx}{0}
\newcommand{\aiiMiCy}{0}
\newcommand{\aiiMiCz}{0}
\newcommand{\aiiMiiCx}{5}
\newcommand{\aiiMiiCy}{0}
\newcommand{\aiiMiiCz}{0}
\newcommand{\aiiMiiiCx}{2.5}
\newcommand{\aiiMiiiCy}{4.3}
\newcommand{\aiiMiiiCz}{0}
\newcommand{\aiiMivCx}{-2.5}
\newcommand{\aiiMivCy}{4.3}
\newcommand{\aiiMivCz}{0}
\newcommand{\aiiMvCx}{-5.0}
\newcommand{\aiiMvCy}{0}
\newcommand{\aiiMvCz}{0}
\newcommand{\aiiMviCx}{-2.5}
\newcommand{\aiiMviCy}{-4.3}
\newcommand{\aiiMviCz}{0}
\newcommand{\aiiMviiCx}{2.5}
\newcommand{\aiiMviiCy}{-4.3}
\newcommand{\aiiMviiCz}{0}
\newcommand{\aiiiMiCx}{0}
\newcommand{\aiiiMiCy}{0}
\newcommand{\aiiiMiCz}{0}
\newcommand{\aiiiMiiCx}{-5}
\newcommand{\aiiiMiiCy}{0}
\newcommand{\aiiiMiiCz}{0}
\newcommand{\aiiiMiiiCx}{5}
\newcommand{\aiiiMiiiCy}{0}
\newcommand{\aiiiMiiiCz}{0}
\newcommand{\aiiiMivCx}{0}
\newcommand{\aiiiMivCy}{-5}
\newcommand{\aiiiMivCz}{0}
\newcommand{\aiiiMvCx}{0}
\newcommand{\aiiiMvCy}{5}
\newcommand{\aiiiMvCz}{0}
\newcommand{\aiiiMviCx}{0}
\newcommand{\aiiiMviCy}{0}
\newcommand{\aiiiMviCz}{-5}
\newcommand{\aiiiMviiCx}{0}
\newcommand{\aiiiMviiCy}{0}
\newcommand{\aiiiMviiCz}{5}

\begin{document}
%
\maketitle
\begin{abstract}
This paper introduces a modification of phase transform on singular value decomposition (SVD-PHAT) to localize multiple sound sources.
This work aims to improve localization accuracy and keeps the algorithm complexity low for real-time applications.
This method relies on multiple scans of the search space, with projection of each  low-dimensional observation onto orthogonal subspaces.
We show that this method localizes multiple sound sources more accurately than discrete SRP-PHAT, with a reduction in the Root Mean Square Error up to 0.0395 radians.
\end{abstract}

\section{Introduction}
\label{sec:introduction}

The cocktail party effect consists of the ability to focus on a specific conversation in a noisy environment.
While humans can usually perform this task efficiently, distant speech processing remains challenging for automatic speech recognition (ASR) systems \cite{tang2018study}.
To improve ASR performances, it is common to use a beamformer with multiple microphones as a preprocessing step to enhance the corrupted speech signal \cite{heymann2015blstm,lee2019deep,sun2018effect}.
Some beamforming methods, such as the delay and sum and the minimum variance distortionless response (MVDR) \cite{habets2010new}, require the target source direction of arrival (DOA).
On the other hand, methods like geometric sound separation require both the target and interference sources direction of arrival \cite{parra2002geometric,valin2004enhanced}.
It is therefore desirable to estimate the direction of arrival of multiple sound sources.

High resolution methods such as Multiple Signal Classification (MUSIC) \cite{schmidt1986multiple} and the Estimation of Signal Parameters via Rotational Invariance Technique (ESPRIT) \cite{roy1986estimation} were initially designed for narrowband signals, and subsequently adapted to broadband signals such as speech \cite{ishi2009evaluation, nakamura2011intelligent, nakamura2012realtime, teutsch2005ebesprit, argentieri2007broadband, danes2010information}.
However, MUSIC-based methods involve online computations of eigenvectors, which makes real-time implementation challenging on low-cost embedded hardware.
On the other hand, ESPRIT-based techniques require significantly less computations, but need twice as many sensors as MUSIC to perform with similar performance, which is problematic for microphone arrays with few sensors.

Alternatively, the Steered-Response Power Phase Transform (SRP-PHAT) robustly estimates the direction of arrival of speech sources and can be computed with low-cost embedded hardware \cite{dibiase2001robust}.
SRP-PHAT relies on the Generalized Cross-Correlation with Phase Transform (GCC-PHAT) between each pair of microphones.
The Fast Fourier Transform is often used to speed up the computation of GCC-PHAT, but this also reduces localization accuracy.
This discrete SRP-PHAT approach can localize many sound sources by scanning the search space multiple times, and nulling the GCC-PHAT region related to each found DOA \cite{grondin2013manyears, valin2007robust, valin2004localization, valin2006robust}.
Hierarchical search also reduces the number of lookups in memory \cite{grondin2019lightweight, yook2016fast}.
The discrete SRP-PHAT approach however relies on rounded TDOA values, which may reduce the localization accuracy.

Alternatively, Cai et al. propose using multiple subbands to individually localize one sound source per band \cite{cai2010localization}.
Similarly, it is possible to localize multiple speech sources based on their distinct pitch values \cite{kepesi2008joint}.
Pavlidi et al. introduce a technique to identify single-source zones in the time-frequency range and generate a histogram to count and localize multiple sound sources \cite{pavlidi2015}.
However, these methods rely on narrow bands to localize sound sources, which makes localization more sensitive to reverberation.
On the other hand, localization can also exploit interesting properties of microphone arrays with symmetrical geometries.
For instance, wavefield decomposition enables localizing multiple sound source with spherical arrays \cite{teutsch2008detection,evers2014multiple,hafezi2016multiple,sun2011robust,nadiri2014localization}.
Similarly, low-complexity multiple sources localization is possible in 2-D with circular arrays \cite{pavlidi2013real,teutsch2006acoustic}.
These methods offer interesting performance, but rely on a specific microphone array geometry, which restricts their scalability.

We recently proposed a new method called SVD-PHAT that relies on singular value decomposition to map the observations to a small subspace, and then uses a nearest neighbor search algorithm like a k-d tree to find the DOA \cite{grondin2019svd}.
This single source localization method is appealing as it preserves exact SRP-PHAT accuracy while greatly reducing the computational complexity, and can adapt to microphone array with arbitrary shapes.
In this paper, we extend SVD-PHAT to localize multiple sound sources.

\section{SRP-PHAT}
\label{sec:srpphat}

We first introduce SRP-PHAT with rounded TDOA that allows efficient localization of multiple sound sources with arbitrary array shapes.
Let $X^l_{m}[k] \in \mathbb{C}$ be the Short Time Fourier Transform (STFT) coefficients, where $N \in \mathbb{N}$ and $\Delta N \in \mathbb{N}$ stand for the frame and hop sizes in samples, respectively, and $k \in \{0,1,\dots,N/2\}$, $m \in \mathcal{M} = \{1,2,\dots,M\}$ and $l \in \mathbb{N}$ stand for the frequency bin, microphone and frame indexes, respectively.
The cross-correlation $X^{l}_{i,j}[k]$ for each microphone pair $(i,j) \in \mathcal{P} = \{(x,y) \in \mathcal{M}^2: x < y\}$ is obtained with the following recursive estimation with $\alpha \in [0,1]$:
\begin{equation}
    X^{l}_{i,j}[k] = (1-\alpha) X^{l-1}_{i,j}[k] + \alpha X^{l}_{i}[k] (X^{l}_{j}[k])^*
\end{equation}
where $\{\dots\}^*$ stands for the complex conjugate.
For clarity, the frame index $l$ is omitted in this paper without loss of generality.
The phase transformed spectrum $\hat{X}_{i,j}[k] \in \mathbb{C}$ for each microphone pair is then obtained in (\ref{eq:srpphat_phat}), where $|\dots|$ stands for the absolute value.

\begin{equation}
    \hat{X}_{i,j}[k] = X_{i,j}[k]/|X_{i,j}[k]|
    \label{eq:srpphat_phat}
\end{equation}

The generalized cross correlation with phase transform (GCC-PHAT) for each pair of microphone and TDOA $\tau \in \mathbb{R}$ is given in (\ref{eq:srpphat_xcorr}), where $W[k,\tau] = \exp{(2\pi\sqrt{-1}k\tau/N)}$.
\begin{equation}
    x_{i,j}[\tau] = \sum_{k=0}^{N/2}{\hat{X}_{i,j}[k]W[k,\tau]}
    \label{eq:srpphat_xcorr}
\end{equation}

The TDOA $\tau_{i,j,q} \in \mathbb{R}$ (in samples) corresponds to the difference between the direction of arrival (DOA) from a source $\mathbf{s}_q \in \{\mathbf{x} \in \mathbb{R}^{3}: \lVert \mathbf{x} \rVert_2 = 1\}$ to microphone $i$ at position $\mathbf{r}_{i} \in \mathbb{R}^{3}$, and the DOA between the same source and another microphone $j$ at position $\mathbf{r}_{j} \in \mathbb{R}^{3}$, scaled with the speed of sound in air $c \in \mathbb{R}^+$ (in m/sec) and the sample rate ($f_S \in \mathbb{N}$):
\begin{equation}
    \tau_{i,j,q} = \frac{f_S}{c}(\mathbf{r}_{j} - \mathbf{r}_{i}) \cdot \mathbf{s}_q
    \label{eq:srpphat_tdoa}
\end{equation}
where $\{\cdot\}$ stands for the dot product.

It is common to discretize $\tau_{i,j,q}$ by rounding to the closest integer (denoted as $\lfloor\tau_{i,j,q}\rceil \in \mathbb{Z}$), and compute the GCC-PHAT in (\ref{eq:srpphat_xcorr}) using an Inverse Fourier Transform (IFFT) for all $\tau = n \in \mathcal{N} = \{0, 1, \dots, N-1\}$.
The expression $Y_{q} \in \mathbb{R}$ is then obtained as follows for every possible DOA $\mathbf{d}_q$, where $q \in \mathcal{Q} = \{1, 2, \dots, Q\}$::
\begin{equation}
    Y_{q} = \sum_{(i,j)\in\mathcal{P}}{x_{i,j}[\lfloor\tau_{i,j,q}\rceil\Mod N]}
    \label{eq:srpphat_Yq}
\end{equation}

The index of the most likely DOA then corresponds to:
\begin{equation}
    q^* = \argmax_{q \in \mathcal{Q}}{\{Y_{q}\}}
    \label{eq:srpphat_q}
\end{equation}

Once the DOA at index $q^*$ is found, we can remove its contribution from the current observations and perform a new scan to detect other active sound sources.
A naive approach consists in nulling the expression $Y_{q^*}$ and some of its closest neighbors, and then scan for a new maximum value.
However, this approach ignores the possible side lobes generated by the found source, and these may lead to false positives in the next scan iteration.
To address this issue, a popular solution consists in nulling some regions in the GCC-PHAT results instead, recompute $Y_q\ \forall\ q \in \mathcal{Q}$ with (\ref{eq:srpphat_Yq}), and then find a new maximum as in (\ref{eq:srpphat_q}).
For each DOA index $q$, we define a subset of DOA indexes $\mathcal{Q}_q = \{x \in \{1, 2, \dots, Q\}: \arccos{(\mathbf{s}_x \cdot \mathbf{s}_q) \leq \Delta\theta}\}$ that gathers DOAs close in space to the DOA $\mathbf{s}_q$, where $\Delta\theta$ is a user-defined parameter that stands for the maximum angle difference.
We then define the set $\mathcal{T}_{i,j,q} = \{\tau_{i,j,q}^{min}, \tau_{i,j,q}^{min}+1, \dots, \tau_{i,j,q}^{max}-1, \tau_{i,j,q}^{max}\}$ that contains all TDOAs related to the DOA $\mathbf{s}_q$ and its closest neighbors, for the microphone pair $(i,j)$, where:
\begin{equation}
\tau_{i,j,q}^{min} = \left\lfloor\min_{p \in \mathcal{Q}_{q}}\{\tau_{i,j,p}\}\right\rfloor
\ \textrm{and}\ 
\tau_{i,j,q}^{max} = \left\lceil\max_{p \in \mathcal{Q}_{q}}\{\tau_{i,j,p}\}\right\rceil
\end{equation}
and the GCC-PHAT values in range of $\mathcal{T}_{i,j,q^*}$ are then set to zero for all pairs.

Algorithm \ref{alg:srpphat} summarizes how SRP-PHAT can be adapted to localize $R$ multiple sources.
At each scan $r \in \mathcal{R} = \{1, 2, \dots, R\}$, the GCC-PHAT values are updated, and the following scans are thus performed without the contribution of the source recently found.
The expressions $\mathbf{d}_r$ and $e_r$ stand for the DOA and energy level found at scan $r$.

\begin{algorithm}[!ht]
    \vspace{4pt}
    \textbf{Offline:}
    \begin{algorithmic}[1]
        \STATE Generate $\tau_{i,j,q}, \mathcal{T}_{i,j,q}\ \forall\ (i,j) \in \mathcal{P}, \forall\ q \in \mathcal{Q}$.
    \end{algorithmic}
    \textbf{Online:}
    \begin{algorithmic}[1]
        \STATE Compute $x_{i,j}[n]\ \forall\ (i,j) \in \mathcal{P}, \forall\ n \in \mathcal{N}$.
        \FOR{$r \in \mathcal{R}$}
        \STATE Compute $Y_q\ \forall\ q \in \mathcal{Q}$.
        \STATE Find $q^*$ using linear search.
        \FOR{$(i,j) \in \mathcal{P}$}
        \STATE $x_{i,j}[\tau] \gets 0, \forall\ \tau \in \mathcal{T}_{i,j,q^*}$
        \ENDFOR
        \STATE $\mathbf{d}_r \gets \mathbf{s}_{q^*}$, $e_r \gets Y_{q^*}$
        \ENDFOR
    \end{algorithmic}    
    \caption{SRP-PHAT for multiple sources}
    \label{alg:srpphat}
\end{algorithm}

Although appealing as it relies on an efficient implementation of GCC-PHAT with IFFTs, this approach relies on discrete cross-correlation results, which reduces the accuracy.
We therefore propose to adapt SVD-PHAT to estimate the direction of arrival (DOA) of multiple sound sources with more accuracy.

\section{SVD-PHAT}
\label{sec:svdphat}

To define the SVD-PHAT method, it is convenient to start from SRP-PHAT in matrix form.
Let us define the vector $\mathbf{X}_{i,j} \in \mathbb{C}^{(N/2+1)\times 1}$ for the microphone pair $(i,j) \in \mathcal{P}$ that holds the phase normalized cross-correlation coefficients for all bins $k \in \{0, 1, \dots, N/2\}$:
\begin{equation}
    \mathbf{X}_{i,j} = \left[
        \begin{array}{cccc}
            \hat{X}_{i,j}[0] & \hat{X}_{i,j}[1] & \cdots & \hat{X}_{i,j}[N/2] \\
        \end{array}
    \right]^T
\end{equation}
where $\{\dots\}^T$ stands for the transpose operator.

A single vector $\mathbf{X} \in \mathbb{C}^{P(N/2+1)\times 1}$ then holds all these vectors:
\begin{equation}
    \mathbf{X} = \left[
        \begin{array}{cccc}
            (\mathbf{X}_{1,2})^T & (\mathbf{X}_{1,3})^T & \cdots & (\mathbf{X}_{M-1,M})^T \\
        \end{array}
    \right]^T
\end{equation}

For each pair of microphones $(i,j)$, all coefficients $W[k,\tau] \in \mathbb{C}$ are concatenated in a matrix $\mathbf{W}_{i,j} \in \mathbb{C}^{Q\times (N/2+1)}$:
\begin{equation}
    \mathbf{W}_{i,j} = \left[
    \begin{array}{ccc}
        W[0,\tau_{1,i,j}] & \cdots & W[N/2,\tau_{1,i,j}] \\
        W[0,\tau_{2,i,j}] & \cdots & W[N/2,\tau_{2,i,j}] \\
        \vdots & \ddots & \vdots \\
        W[0,\tau_{Q,i,j}] & \cdots & W[N/2,\tau_{Q,i,j}] \\
    \end{array}
    \right]
\end{equation}

The supermatrix $\mathbf{W} \in \mathbb{C}^{Q\times P(N/2+1)}$ then holds all the matrices $\mathbf{W}_{i,j}\ \forall\ (i,j) \in \mathcal{P}$:
\begin{equation}
\mathbf{W} = \left[
\begin{array}{cccc}
    \mathbf{W}_{1,2} & \mathbf{W}_{1,3} & \cdots & \mathbf{W}_{M-1,M}
\end{array}
\right]
\end{equation}

Finally, the vector $\mathbf{Y} \in \mathbb{R}^{Q\times 1}$ holds the results $Y_q\ \forall\ q \in \{1, 2, \dots, Q\}$, where $\Re\{\dots\}$ returns the real part:
\begin{equation}
    \mathbf{Y} = \left[
        \begin{array}{ccc}
            Y_1 & \dots & Y_Q
        \end{array}
    \right]^T = \Re\{\mathbf{W}\mathbf{X}\}
\end{equation}

The supermatrix $\mathbf{W}$ can be estimated with SVD of rank $K \in \{1, 2, \dots, K_{max}\}$, with $K_{max} = \max\{Q,P(N/2+1)\}$ and where $\mathbf{U} \in \mathbb{C}^{Q\times K}$, $\mathbf{S} \in \mathbb{C}^{K\times K}$ and
$\mathbf{V} \in \mathbb{C}^{P(N/2+1)\times K}$:
\begin{equation}
    \mathbf{W} \approx  \mathbf{U}\mathbf{S}\mathbf{V}^H
\end{equation}

The rank $K$ corresponds to the minimum value for which the following condition holds, where $\delta \in (0,1)$ is a user-defined parameter that stands for the reconstruction tolerable error:
\begin{equation}
    \Tr{\{\mathbf{S}\mathbf{S}^T\}} \geq (1 - \delta)\Tr{\{\mathbf{W}\mathbf{W}^H\}}
    \label{eq:svdphat_cond1}
\end{equation}
where $\Tr\{\dots\}$ stands for the trace of the matrix.

The vector $\mathbf{Z} \in \mathbb{C}^{K \times 1}$ then results from the projection of the observations $\mathbf{X}$ in the $K$-dimensions subpace:
\begin{equation}
    \mathbf{Z} = \mathbf{V}^H\mathbf{X}
\end{equation}

Similarly, we define the dictionary $\mathbf{D} \in \mathbb{C}^{Q \times K}$, made of the vectors $\mathbf{D}_q \in \mathbb{C}^{1 \times K}\ \forall\ q \in \{1, 2, \dots, Q\}$:
\begin{equation}
    \mathbf{D} = \mathbf{U}\mathbf{S} = \left[
    \begin{array}{cccc}
        (\mathbf{D}_1)^T & (\mathbf{D}_2)^T & \dots & (\mathbf{D}_Q)^T \\
    \end{array}
    \right]^T
\end{equation}

As explained in \cite{grondin2019svd}, the DOA index then corresponds to $q^*$, obtained as follows:
\begin{equation}
    q^* = \argmax_{q \in \mathcal{Q}}{\{\Re\{\mathbf{D}_q \cdot \mathbf{Z}^H\}\}}
\end{equation}
which can be converted into the following nearest neighbor problem with an algorithm such as k-d tree:
\begin{equation}
    q^* = \argmin_{q \in \mathcal{Q}}{\{\lVert\hat{\mathbf{D}}_{q}-\hat{\mathbf{Z}}^H\rVert_2^2\}}
\end{equation}
where $\hat{\mathbf{D}}_q = \mathbf{D}_q / \lVert \mathbf{D}_q \rVert_2$ and $\hat{\mathbf{Z}} = \mathbf{Z} / \lVert \mathbf{Z} \rVert_2$.

Intuitively, we would like to remove the component in $\mathbf{Z}$ that spans the space spanned by $(\mathbf{D}_{q^*})^*$, and then perform a new scan to find another source.
We thus define the vector $\mathbf{v}_r \in \mathbb{C}^{1 \times K}$ as follows:
\begin{equation}
    \mathbf{v}_r = (\mathbf{D}_{q^*})^*
\end{equation}

The Gram-Schmidt process then makes the current vector $\mathbf{v}_r$ at scan $r$ orthogonal to all the vectors previously found ($\hat{\mathbf{u}}_n\ \forall\ n \in \{1,2,\dots,r-1\}$), and generates $\mathbf{u}_r$:
\begin{equation}
    \mathbf{u}_r = \mathbf{v}_r - \sum_{n=1}^{r-1}{(\hat{\mathbf{u}}_n \cdot \mathbf{v}_r)\hat{\mathbf{u}}_n}
    \label{eq:svdphat_gs}
\end{equation}
which is then normalize to have a unit norm:
\begin{equation}
    \hat{\mathbf{u}}_r =\mathbf{u}_r/\lVert \mathbf{u}_r \rVert_2
\end{equation}

Finally, the current observation $\mathbf{Z}$ is projected in the subspace orthogonal to $\hat{\mathbf{u}}_r$ to remove the current contribution of the source previously found:
\begin{equation}
    \mathbf{Z}' = \mathbf{Z} - (\hat{\mathbf{u}}_r \cdot \mathbf{Z})\hat{\mathbf{u}}_r
\end{equation}

Algorithm $\ref{alg:svdphat}$ summarizes these steps for SVD-PHAT.
This approach is appealing as it involves $R$ k-d tree search instead of computing $R$ times $Y_q\ \forall\ q \in \mathcal{Q}$ as in (\ref{eq:srpphat_Yq}), which reduces the algorithm complexity.

\begin{algorithm}[!ht]
    \vspace{4pt}
    \textbf{Offline:}
    \begin{algorithmic}[1]
        \STATE Generate $\mathbf{D}$, $\mathbf{V}$, and $\bar{\mathbf{V}}_q\ \forall\ q \in \{1, 2, \dots, Q\}$.
    \end{algorithmic}
    \textbf{Online:}
    \begin{algorithmic}[1]
        \STATE Compute $\mathbf{Z}$ from $\mathbf{V}$ and observations $\mathbf{X}$.
        \FOR{$r \in \{1, 2, \dots, R\}$}
        \STATE Find $q^*$ using a k-d tree to minimize $\lVert\hat{\mathbf{D}}_{q}-\hat{\mathbf{Z}}^H\rVert_2^2$.
        \STATE Compute $Y_{q^*}$, $\mathbf{v}_r$, $\hat{\mathbf{u}}_r$ and $\mathbf{Z}'$.
        \STATE $\mathbf{Z} \gets \mathbf{Z}'$, $\mathbf{d}_r \gets \mathbf{s}_{q^*}$, $e_r \gets Y_{q^*}$
        \ENDFOR
    \end{algorithmic}    
    \caption{SVD-PHAT for multiple sources}
    \label{alg:svdphat}
\end{algorithm}

\section{RESULTS}
\label{sec:results}

We investigate three different microphone array geometries: a 1-D linear array, a 2-D planar array and a 3-D array.
The microphones xyz-positions with respect to the center of the array are given in cm in Table \ref{tab:results_positions}.

\begin{table}[!ht]
    \centering
    \caption{Positions (x,y,z) of the microphones in cm}
    \renewcommand{\arraystretch}{1.2}
    \begin{tabular}{|c|c|c|c|}
    \hline
    Mic & 1-D & 2-D & 3-D \\
    \hline
    $1$ & $(\aiMiCx,\aiMiCy,\aiMiCz)$ 
        & $(\aiiMiCx,\aiiMiCy,\aiiMiCz)$ 
        & $(\aiiiMiCx,\aiiiMiCy,\aiiiMiCz)$ \\
    $2$ & $(\aiMiiCx,\aiMiiCy,\aiMiiCz)$ 
        & $(\aiiMiiCx,\aiiMiiCy,\aiiMiiCz)$ 
        & $(\aiiiMiiCx,\aiiiMiiCy,\aiiiMiiCz)$ \\
    $3$ & $(\aiMiiiCx,\aiMiiiCy,\aiMiiiCz)$ 
        & $(\aiiMiiiCx,\aiiMiiiCy,\aiiMiiiCz)$ 
        & $(\aiiiMiiiCx,\aiiiMiiiCy,\aiiiMiiiCz)$ \\
    $4$ & $(\aiMivCx,\aiMivCy,\aiMivCz)$ 
        & $(\aiiMivCx,\aiiMivCy,\aiiMivCz)$ 
        & $(\aiiiMivCx,\aiiiMivCy,\aiiiMivCz)$ \\
    $5$ & $(\aiMvCx,\aiMvCy,\aiMvCz)$ 
        & $(\aiiMvCx,\aiiMvCy,\aiiMvCz)$ 
        & $(\aiiiMvCx,\aiiiMvCy,\aiiiMvCz)$ \\
    $6$ & $(\aiMviCx,\aiMviCy,\aiMviCz)$ 
        & $(\aiiMviCx,\aiiMviCy,\aiiMviCz)$ 
        & $(\aiiiMviCx,\aiiiMviCy,\aiiiMviCz)$ \\
    $7$ & $(\aiMviiCx,\aiMviiCy,\aiMviiCz)$
        & $(\aiiMviiCx,\aiiMviiCy,\aiiMviiCz)$ 
        & $(\aiiiMviiCx,\aiiiMviiCy,\aiiiMviiCz)$ \\
    \hline
    \end{tabular}
    \label{tab:results_positions}
\end{table}

Simulations are conducted to measure the accuracy of the proposed method and compare it to the SRP-PHAT approach discretized with IFFTs.
The microphone array is positioned and rotated randomly in a $10\textrm{m} \times 10\textrm{m} \times 3\textrm{m}$ rectangular room, with a minimum distance of $0.5\textrm{m}$ from the walls, ceiling and floor.
The target sources are also positioned randomly in the room, and the random setup ensures a minimum angle difference of $30^{\circ}$ between each source, a distance of at least $0.5\textrm{m}$ between each source and the center of the microphone array, and a distance of at least $0.5\textrm{m}$ between each source and the walls, ceiling and floor.
For each configuration, the room reverberation is modeled with Room Impulse Responses (RIRs) generated with the image method \cite{allen1979image}, where the reverberation time (RT60) is sampled randomly in the uniform interval between $200$ and $500$ msecs, which corresponds to the levels previously used in \cite{grondin2019svd}.
Sound segments selected randomly from the TIMIT dataset \cite{zue1990speech} are normalized to have the same energy levels, and are convolved with the generated RIRs.
For each type of array (1-D, 2-D and 3-D) and number of active sources (1, 2 and 3), we perform $1000$ simulations.

Table \ref{tab:results_parameters} introduces the parameters used with SRP-PHAT and SVD-PHAT.
The sample rate $f_S$ captures all the frequency content of speech (including wideband fricatives that contain relevant localization information), and the speed of sound $c$ matches typical indoor conditions at room temperature.
The frame size $N$ analyze speech segments of $32$ msecs, and the hop size provides a $50\%$ overlap.
The DOAs are scanned on a unit sphere generated recursively from a tetrahedron, for a total of $2562$ points, as in \cite{grondin2019lightweight}.
Moreover, the cross-correlation adaptation rate $\alpha$ estimates the sound statistic over the past $400$ msecs.
In the case of SRP-PHAT, the maximum angle difference $\Delta\theta$ corresponds to $0.1745\ \textrm{radians}$ to null the current source within a region of $10^{\circ}$.
In the specific case of SVD-PHAT, the user-defined parameter is set to $\delta = 10^{-5}$ as in \cite{grondin2019svd}.

\begin{table}[!ht]
    \centering
    \caption{Parameters for SRP-PHAT and SVD-PHAT}
    \renewcommand{\arraystretch}{1.2}
    \begin{tabular}{|cccccc|}
        \hline
        $f_S$ & $c$ & $N$ & $\Delta N$ & $Q$ & $\alpha$ \\
        \hline
        $16000$ & $340.0$ & $512$ & $128$ & $2562$ & $0.1$ \\
        \hline
    \end{tabular}
    \label{tab:results_parameters}
\end{table}

For a 1-D array with all microphones on the x-axis, the spatial resolution is limited to an arc that goes from $0^{\circ}$ to $180^{\circ}$ in the xy-plane.
The 3-D DOA is therefore projected to this subspace as follows:
\begin{equation}
    f_1(\mathbf{x}) = [\cos(g(\mathbf{x})), \sin(g(\mathbf{x})), 0]
\end{equation}
where:
\begin{equation}
    g(\mathbf{x}) = \atantwo\left\{(\mathbf{x})_x,\sqrt{(\mathbf{x})_y^2+(\mathbf{x})_z^2}\right\}
    \label{eq:results_gx}
\end{equation}

Similarly, a 2-D array that spans the xy-plane allows DOA estimation on a half hemisphere oriented in the z-axis, and therefore all DOAs are projected to the positive z-axis:
\begin{equation}
    f_2(\mathbf{x}) = [(\mathbf{x})_x, (\mathbf{x})_y, |(\mathbf{x})_z|]
\end{equation}

Finally, for a 3-D array, the DOA can span the full space:
\begin{equation}
    f_3(\mathbf{x}) = \mathbf{x}
\end{equation}

For each speech source $t\ \forall\ \{1,2,\dots,T\}$, the minimum angle difference (in $\textrm{radians}$) between the theoretical DOA and all estimated DOAs at frame $l \in \{1, 2, \dots, L\}$ is given as follows:
\begin{equation}
    \phi_{t}^l = \min_{r \in \mathcal{R}} \left\{\arccos{(f_{\beta}(\mathbf{d}^l_r) \cdot f_{\beta}(\mathbf{c}_t))}\right\}
\end{equation}
where $\beta \in \{1,2,3\}$ matches the array geometry.
The goal is therefore to have at least one DOA estimation that matches each speech source true DOA.

In the proposed experiments, the number of sources varies with $T = \{1,2,3\}$, and the number of scans $R$ matches this number.
The root mean square error (RMSE) in rad for a simulation therefore corresponds to:
\begin{equation}
    RMSE = \sqrt{\frac{1}{LT}\sum_{l=1}^{L}\sum_{t=1}^{T}{(\phi^l_t)^2}}
\end{equation}

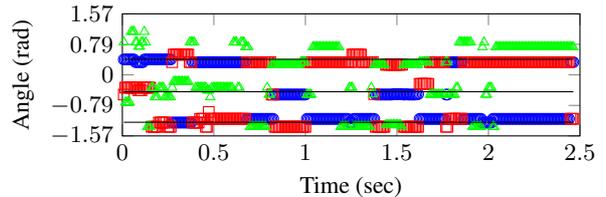
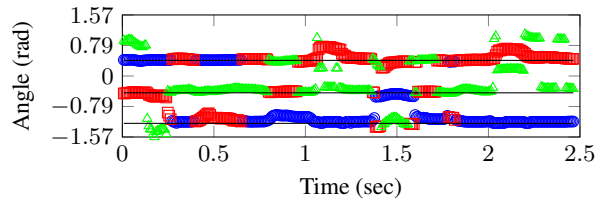
\begin{figure}[!ht]
    \centering
    \subfloat[SRP-PHAT (RMSE = 0.3009)]{%
        \begin{tikzpicture}
        \begin{axis}[xlabel=Time (sec),ylabel=Angle (rad),ymin=-1.5708,ymax=1.5708,ytick={-1.5708,-0.7854,0,0.7854,1.5708},xmin=0,xmax=2.5,clip mode=individual, height=0.4\columnwidth,width=0.95\columnwidth]
        \addplot[blue,only marks,mark=o] table [x=time, y=r1, col sep=comma] {data/results_srp.csv};
        \addplot[red,only marks,mark=square] table [x=time, y=r2, col sep=comma] {data/results_srp.csv};        
        \addplot[green,only marks,mark=triangle] table [x=time, y=r3, col sep=comma] {data/results_srp.csv};
        \addplot[black] table [x=time, y=s1, col sep=comma] {data/results_srp.csv};
        \addplot[black] table [x=time, y=s2, col sep=comma] {data/results_srp.csv};
        \addplot[black] table [x=time, y=s3, col sep=comma] {data/results_srp.csv};
        \end{axis}
        \end{tikzpicture}} \\
    \subfloat[SVD-PHAT (RMSE = 0.2027)]{%
        \begin{tikzpicture}
        \begin{axis}[xlabel=Time (sec),ylabel=Angle (rad),ymin=-1.5708,ymax=1.5708,ytick={-1.5708,-0.7854,0,0.7854,1.5708},xmin=0,xmax=2.5,clip mode=individual,height=0.4\columnwidth,width=0.95\columnwidth]
        \addplot[blue,only marks,mark=o] table [x=time, y=r1, col sep=comma] {data/results_svd.csv};
        \addplot[red,only marks,mark=square] table [x=time, y=r2, col sep=comma] {data/results_svd.csv};        
        \addplot[green,only marks,mark=triangle] table [x=time, y=r3, col sep=comma] {data/results_svd.csv};
        \addplot[black] table [x=time, y=s1, col sep=comma] {data/results_svd.csv};
        \addplot[black] table [x=time, y=s2, col sep=comma] {data/results_svd.csv};
        \addplot[black] table [x=time, y=s3, col sep=comma] {data/results_svd.csv};
        \end{axis}
        \end{tikzpicture}}     
        \caption{Azimuth angle (in rad, obtained with $g(\mathbf{x})$ in (\ref{eq:results_gx})) of potential sources found with SRP-PHAT and SVD-PHAT, for $r=1$ (blue circles), $r=2$ (red squares) and $r=3$ (green triangles). The theoretical DOAs are $-1.2192$, $-0.4335$ and $0.4015$, and are plotted with solid black lines. For this simulation, the reverberation time (RT60) corresponds to 238 msecs.}
    \label{fig:results_doas}
\end{figure}

Figure \ref{fig:results_doas} shows the estimated DOAs obtained with SRP-PHAT and SVD-PHAT for a 1-D array with three speech sources located at $-1.2192$ rad, $-0.4335$ rad and $0.4015$ rad, and a reverberation time (RT60) of $238$ msecs.
In this example, the SRP-PHAT method fails to detect the source at $-0.4335$ rad at different times, whereas SVD-PHAT detects this source most of the time.
The RMSEs of SRP-PHAT and SVD-PHAT correspond to $0.3009$ rad and $0.2027$ rad, respectively, which indicates that SVD-PHAT outperforms SRP-PHAT in this specific example.

The RMSE gap between both SRP-PHAT and SVD-PHAT is however usually smaller than the one shown in Figure \ref{fig:results_doas}.
To better compare both methods, Table \ref{tab:results_rmse} shows the mean of all RMSEs for the $1000$ simulations with each configuration.
In all cases, the proposed multiple source SVD-PHAT reduces the RMSE compared to the discrete SRP-PHAT, but with a smaller gap that oscillates between $0.0244$ rad and $0.0395$ rad.
It is interesting to note that for multiple sound sources ($T>1$), the RMSE increases rapidly.
This is expected as multiple active sources partially overlap each other in the time-frequency domain, which makes localization more challenging.
The best improvement for multiple sources occurs for the 3-D array when two sources are active, with a reduction in the RMSE of $0.0395$ rad.

\begin{table}[!ht]
    \centering
    \caption{Root Mean Square Error (RMSE) -- less is better}
    \renewcommand{\arraystretch}{1.2}
    \begin{tabular}{|c|c|c|c|c|}
        \hline
        Geometry & Nb. Sources & SRP-PHAT & SVD-PHAT \\
        \hline
        \multirow{3}{*}{1-D} & $1$ & $0.0884$ & $\mathbf{0.0509}$ \\
                             & $2$ & $0.2656$ & $\mathbf{0.2274}$ \\
                             & $3$ & $0.2763$ & $\mathbf{0.2519}$ \\
        \hline
        \multirow{3}{*}{2-D} & $1$ & $0.1356$ & $\mathbf{0.0820}$ \\
                             & $2$ & $0.4516$ & $\mathbf{0.4200}$ \\
                             & $3$ & $0.4201$ & $\mathbf{0.3828}$ \\
        \hline
        \multirow{3}{*}{3-D} & $1$ & $0.0708$ & $\mathbf{0.0296}$ \\
                             & $2$ & $0.4550$ & $\mathbf{0.4155}$ \\
                             & $3$ & $0.5445$ & $\mathbf{0.5189}$ \\
        \hline
    \end{tabular}
    \label{tab:results_rmse}
\end{table}

\section{CONCLUSION}
\label{sec:conclusion}

This paper extends SVD-PHAT for multiple sound source localization.
This technique outperforms the discrete SRP-PHAT approach in terms of accuracy, while preserving the low complexity of the original SVD-PHAT.
On average, the reduction in the RMSE varies between $0.0244$ and $0.0395$ radians, and the best improvement is observed for an array that spans 3-D space with two simultaneous speech sources.

In future work, we will investigate alternatives to k-d tree search to address the curse of dimensionality during the nearest neighbor search \cite{berchtold1998pyramid}.
The method could also be extended to deal with speed of sound mismatch and the near-field effect.
Microphone directivity could also be combined with SVD-PHAT to make the propagation model more realistic \cite{thomas2012optimal}.
The sound source tracking method proposed in \cite{grondin2019lightweight} could also be combined to SVD-PHAT to estimate the number of sound sources and track their positions over time.
Finally, it would be interesting to implement SVD-PHAT in C code for easy deployment on real-time embedded systems.

\vfill\pagebreak

\bibliographystyle{IEEEtran}
\bibliography{refs}

\end{document}